*Short Note on Superconductivity at Ambient Temperature and Pressure in Silver Embedded Gold Nano-particles: A Goldsmith job ahead*

*V.P.S. Awana*

*National Physical laboratory, Dr. K.S. Krishnan Marg, New Delhi-110012, India*

Very recent observation of superconductivity (arxiv: 1807.08572) in both transport (resistance versus temperature) and magnetization (Zero field cooled) at around above 235K has obviously raised many eyebrows. The experimentalists in particular are more thrilled than the theoreticians. Although the results presented in the paper are clean and both transport and magnetization measurements do approve the observation of superconductivity to some extent (the dia-magnetic shielding fraction is low), the problem is to synthesize/fabricate the material. This I name a "goldsmith" job, primarily because one has to deal with Gold and Silver *nano*-particles combination/amalgamation/alloying or even who knows the interfacial new phase, which could be superconducting. This short note is written in view to let the spark continues and the interesting work of (arxiv: 1807.08572) authors is reproduced independently.

Superconductivity being often quoted as "Superconductivity 100 Years young" seems to have been getting even younger by the day. Thanks to a very recent report which appeared around two weeks before on arxiv claiming superconductivity at ambient temperature and pressure in Gold-Silver *nano*-particles [1]. Keeping aside may be very few exceptions, the phenomenon of superconductivity [2] along with the characteristics of most of superconducting materials were well explained or predicted on the basis of popular BCS (Bardeen Cooper Schrieffer) theory [3]. The big surprise happened in 1986, when superconductivity was observed in Cuprates at above 30K [4], which got raised to above 90K in short span of time [5]. Later the Cuprates superconductivity raised to as high as in excess of 130K in HgBaCaCuO layered compounds [6]. The 1986 discovery of appearance of superconductivity to the tune of above 130K in Cuprates is one of the most happening scientific events of the last century. Later superconductivity in excess of 20K was reported in quaternary Boro-Carbides (RENi$_2$B$_2$C) along with its interesting interplay with magnetism [7].

Theoreticians are yet struggling to find a unified theory to explain high T$_c$ superconductivity of Cuprates. More latter leaving aside the discovery of superconductivity in MgB$_2$ at 39K in 2001 from J. Akimitsu group in Japan [8], which was yet explainable in strong BCS limit, nothing very exciting did happen in field of supercondtivity. However, thunder struck again in 2008, when superconductivity was discovered by H. Hosono group in Japan in excess of 30K in Layered FaAs based LaFeAsO/F [9]. Interestingly, the FeAs based layered compounds superconductivity did soon jumped to above 55K in case of SmFeAsO/F [10] and thus surpassing the popular BCS limit. Soon after, superconductivity was seen in iron chalcegonide (FeSe) compounds [11], which increased to above 100K under pressure or tensile stress [12]. More recently, superconductivity is seen in sulfur hydride at 203K under pressure of above 150GPa [13]. Summarily, in a layman language the superconductivity story from discovery [2] to BCS theory [3] to Cuprates [4-6], Iron based compounds [9-12], and sulfur hydride [13] has been very fascinating for the scientific community. Moreover, there have been various theoretical advancements as well, along with the state of art experimental probes are been used to unearth the mystery of superconductivity beyond BCS.

The latest discovery [1], i.e., superconductivity at ambient pressure and temperature in Gold-Silver *nano* particles will hopefully take the story further. As far as the synthesis of the material is concerned, let us honestly ask ourselves how many of us could believe a few decades before about the high surface conductivity of topological insulators [14], or for that matter the Graphene [15] and even the recent 2D wonder materials viz. $MoS_2$ etc. [16]. The important point about the recent report [1] is that here one gets high temperature superconductivity among or within the possibly interacting individually non superconducting allies down to lowest temperatures. Further, the control of size i.e., *nano*-structuring of a superconductor generally results in suppression of superconductivity. This makes the present discovery [1] further interesting or may be intriguing due to the fact that whether the observed superconductivity is of the interacting *nano*-particles or otherwise. In this direction one gets remind of the fact the superconductivity does appear at the interface of two insulators i.e., $SrTiO_3$/$LaTiO_3$ [17]. In present case [1], what if superconductivity is at (a) interface of *nano* Gold and Silver particles, (b) of granular nature through the alloyed Gold-Silver composite or (c) the new phase boundary structure giving rise to quantum state viz. topological insulator edge states. Remember in present case [1] the Silver particles are embedded in Gold matrix.

Any case, as far as the exact location (interfaces/grain boundary/edge states) of the superconductivity is concerned the same could be looked after from various local probes or transport and spectroscopy measurements, only after one gets through the reproducible quality material in question. In this direction, the strategy one could be to follow the process of the authors [1] and second be to deal with other novel ways to get Silver embedded Gold superconductor. I would suggest preferring both and in fact rather more attention to the latter option. For example,

(a) Have Gold and Silver *nano*-particles synthesized separately, mix them gently in different ratio, press in regular shape and bake at moderate temperature and characterize the resultant material for transport and magnetization in first run. The mixing ratio and baking temperature are important parameters, who know at what combinations the desired surface quantum state does exist. The embedment of the two (Silver and Gold) *nano*-particles is important here.

(b) Second strategy could be to grow the Gold and silver alloyed structures from their liquid state i.e. after vacuum heating them at above 1000 $^0$C and cooled slowly. Various combinations related to different stoichiometric ratios and growth parameters will be required. This could be a situation like $Bi_{1-x}Sb_x$ [18]. The resultant sample must every time be subjected to transport and magnetization measurements.

(c) Third strategy may include the controlled interfaces by evaporation based thin film heterostructures of Gold and Silver, which as such may not be the straight forward task.

The author would further like to state that strategy a, b or c may not be sufficient or fruitful at all, however one has to start from somewhere and secondly the same would pave the way for new avenues. The claimed discovery of superconductivity [1] needs the approval of scientific community. The results shown in ref. [1], exhibiting clear transport and magnetization signatures of superconductivity at around 235K bounds one to think positively and to try to get the new wonder superconductor. The proper *nano*-structuring of Gold-Silver combination could be the right key. It is really the Goldsmith job and that also in a more precise scientific way. It seems one day the *jewelers will be selling the superconducting jewelry, i.e., Silver embedded Gold superconductor.*

On theoretical front already an interesting article related to the existence of confined High $T_c$ superconductivity phase, behind a robust Fermi liquid in monovalent metals has just appeared on arxiv [18]. One of the interesting suggested experiments would be to perform the Au isotope effect for new superconductor [1], to facilitate in knowing whether BCS is effective here or not [3]. In summary, the current short note is written with sole aim to spark interest in *nano*-structured Gold-Silver superconductivity of 225K. Further few suggestions are put forward to synthesize/fabricate the new possible wonder superconductor.